\documentclass[trackchanges]{aastex7}
\usepackage[utf8]{inputenc}
\usepackage{ragged2e}

\usepackage{natbib}
\usepackage{textcomp} 
\begin{document}

\title{A Catalogue of Mid-infrared Variable Sources from unTimely}

\author[orcid=0000-0003-3778-3566, gname=Zihan, sname='Kang']{Zihan Kang}
\affiliation{National Astronomical Observatories, Chinese Academy of Sciences, Beijing 100101, People's Republic of China}
\affiliation{University of Chinese Academy of Science, Beijing 100049, People's Republic of China}
\email{zhkang@bao.ac.cn}

\author[orcid=0000-0002-7048-0930, gname=Jingyi, sname='Zhang']{Jingyi Zhang}
\affiliation{National Astronomical Observatories, Chinese Academy of Sciences, Beijing 100101, People's Republic of China}
\email{jyzhang@bao.ac.cn}  

\author[orcid=0000-0002-6610-5265, gname=Yanxia, sname='Zhang']{Yanxia Zhang} 
\affiliation{National Astronomical Observatories, Chinese Academy of Sciences, Beijing 100101, People's Republic of China}
\email{zyx@bao.ac.cn}

\author[gname=Changhua, sname='Li']{Changhua Li}
\affiliation{National Astronomical Observatories, Chinese Academy of Sciences, Beijing 100101, People's Republic of China}
\affiliation{National Astronomical Data Center, Beijing 100101, People's Republic of China}
\email{lich@bao.ac.cn}

\author[gname=Xiao, sname='Kong']{Xiao Kong}
\affiliation{National Astronomical Observatories, Chinese Academy of Sciences, Beijing 100101, People's Republic of China}
\email{kongx@bao.ac.cn}

\author[gname=Minzhi, sname='Kong']{Minzhi Kong}
\affiliation{College of Physics, Hebei Normal University, Shijiazhuang 050024, People's Republic of China}
\affiliation{Guo Shoujing Institute for Astronomy, Hebei Normal University, Shijiazhuang 050024, People’s Republic of China}
\email{kmz@hebtu.edu.cn}

\author[gname=Jinghang, sname='Shi']{Jinghang Shi}
\affiliation{National Astronomical Observatories, Chinese Academy of Sciences, Beijing 100101, People's Republic of China}
\affiliation{University of Chinese Academy of Science, Beijing 100049, People's Republic of China}
\email{shijinghang24@mails.ucas.ac.cn}

\author[gname=Shirui, sname='Wei']{Shirui Wei}
\affiliation{National Astronomical Observatories, Chinese Academy of Sciences, Beijing 100101, People's Republic of China}
\affiliation{University of Chinese Academy of Science, Beijing 100049, People's Republic of China}
\email{weisr@bao.ac.cn}

\author[0000-0002-7350-6913]{Xue-Bing Wu}
\email{wuxb@pku.edu.cn}
\affil{National Astronomical Observatories, Chinese Academy of Sciences, Beijing 100101, People's Republic of China}
\affil{Department of Astronomy, School of Physics, Peking University, Beijing 100871, People's Republic of China}
\affil{Kavli Institute for Astronomy and Astrophysics, Peking University, Beijing 100871, People's Republic of China}

\correspondingauthor{Yanxia Zhang}
\email{zyx@bao.ac.cn}
\correspondingauthor{Jingyi Zhang}
\email{jyzhang@bao.ac.cn}
\correspondingauthor{Xiao Kong}
\email{kongx@bao.ac.cn}

\begin{abstract}
The WISE and NEOWISE missions have provided the only mid-infrared all-sky time-domain data, opening a unique observational window for variability studies. Yet, a comprehensive and systematic catalog of mid-infrared variable sources has remained unavailable. In this work, we construct the first large-scale mid-infrared variability catalog based on the unTimely coadded photometry, covering tens of millions of sources. By employing a Bayesian Gaussian mixture model with a Dirichlet process, we identified 8,256,042 variable sources in the W1 band and 7,147,661 in the W2 band, significantly expanding the landscape of known mid-infrared variables. In addition to robust variability metrics, our analysis highlights rare and extreme outliers through dedicated outlier-detection algorithms, enabling the discovery of unusual classes of objects such as eruptive young stellar objects, highly variable active galactic nuclei, and other rare transients. This unprecedented dataset provides a new foundation for time-domain astronomy in the mid-infrared, offering complementary insights to optical and near-infrared surveys, and opening the door to systematic investigations of stellar evolution, accretion processes, and dust-enshrouded astrophysical environments on a Galactic and extragalactic scale.


\end{abstract}



\section{Introduction}

As we enter the third decade of the 21st century, time-domain astronomy has gained significant prominence. This field enables the examination of a rapidly evolving universe on timescales perceptible to human experience. Various ongoing and forthcoming large-scale sky surveys are transforming our understanding of astronomical objects, including the internal structures of stars, the physical processes in accretion disks, and transient phenomena. However, due to the limitations of the atmospheric window, most surveys are conducted in the optical band, leaving other wavelength ranges insufficiently explored. One crucial regime is the mid-infrared, as such radiation can penetrate dust to reveal obscured objects and probe dust properties. Furthermore, many cold celestial bodies have radiation peaks in the mid-infrared band, making it essential to study their properties in this band.

To date, the NEOWISE \citep{2014ApJ...792...30M} project of the WISE satellite represents the sole all-sky survey conducted at mid-infrared wavelength. The Wide-field Infrared Survey Explorer (WISE; \citealt{2010AJ....140.1868W}) is a NASA-operated infrared-wavelength astronomical space telescope that was launched in December 2009. Its primary mission, executed in 2010, involved conducting an all-sky survey across four mid-infrared bands (W1-W4), resulting in an exceptionally deep and comprehensive catalog of celestial objects. Following the depletion of its cryogenic coolant, two of the four bands (W1 at 3.4 $\mu$m and W2 at 4.6 $\mu$m) remained operational. The mission was subsequently reinitiated as NEOWISE in 2013, focusing on the discovery and characterization of near-Earth objects (NEOs). However, through its repeated surveys of the entire sky over more than a decade, the NEOWISE reactivation mission has serendipitously produced an unparalleled, long-term temporal dataset for hundreds of millions of astronomical sources at mid-infrared wavelengths. This extensive archive offers a unique opportunity to systematically study the intrinsic variability across the sky.

Numerous catalogues have been developed to integrate the WISE data. In addition to the official database, resources such as unWISE \citep{2019ApJS..240...30S} and catWISE \citep{2021ApJS..253....8M} provide valuable supplements. However, over the years, the utilization of NEOWISE variable sources' data has predominantly been limited to referencing specific sources of interest. The lack of systematic selection of mid-infrared variable sources has hindered studies based on comprehensive sky statistics and the identification of special variable candidates. Except the large level of error of the photometry, a significant reason for this limitation may be the difficulty in accessing the entire database for the analysis. Fortunately, the unTimely catalogue \citep{2023AJ....165...36M} offers an opportunity to conveniently access all variables with long time scales. Because the WISE satellite conducts observations of most sky regions during two distinct epochs within a year, each lasting approximately two weeks, the unTimely catalogue aggregates coadded images for each epoch and extracts sources to construct a comprehensive photometric catalogue on an annual scale.

In this study, we implemented a cross-match operation to construct light curves for all sky sources in the unTimely catalogue and applied machine learning algorithms to identify variable sources, thereby forming a catalogue and highlighting potentially interesting special variables. This paper is organized as follows: Section \ref{sec:unTimely Catalog} introduces the unTimely catalogue. Section \ref{sec:light curves} describes the construction of the light curves. Section \ref{sec:features} outlines the variability features utilized by the algorithm to identify the variables. The algorithm, named the Bayesian Gaussian mixture model with a Dirichlet process, is detailed in Section \ref{sec:Gaussian mixture}. Subsequently, Section \ref{sec:sample} describes the resultant variable source sample, and Section \ref{sec:Outliers} presents our method for identifying potential special sources. A brief discussion of the potential scientific inquiries that can be pursued using our catalogue is provided in Section \ref{sec:discussion}.

\section{unTimely Catalog} \label{sec:unTimely Catalog} 
The unTimely catalog is derived from WISE and NEOWISE observations spanning 2010-2020. The catalog was extracted from the `time-resolved unWISE coadds' \citep{2018AJ....156...69M}, includes data from the W1 and W2 bands, providing approximately 32 full-sky unWISE catalogs. The catalog offers enhanced depth compared to the WISE/NEOWISE Single Exposure Source Tables, achieving approximately 1.3 magnitudes deeper depth near the ecliptic, with a further improved depth at higher ecliptic latitudes.

The total data volume of the untimely catalog is approximately 4TB, small enough to be stored on a conventional personal hard-disk drive. This relatively manageable data volume renders it accessible to individual researchers, potentially facilitating a comprehensive investigation of full-sky infrared-variable sources.

It is important to note that the unTimely catalog employs a different photometry method compared to the WISE/NEOWISE Single Exposure Source Tables. By implementing a more aggressive deblending strategy, the catalog has the potential to identify additional sources. However, this approach may also complicate the identification of variable sources owing to incorrect deblending. For example, a point-like source may appear as a single object in most epochs but split into multiple detections in others due to observational or processing limitations, creating spurious variability. The situation becomes more complex with regard to extended sources. This presents challenges in selecting true variables with completeness and reliability.

\section{Construction of Light Curves} \label{sec:light curves}
Constructing light curves requires cross-matching numerous full-sky tables, a task that poses significant challenges. However, the unTimely catalog employs the same HEALPix partition of the sky as the ALLWISE catalog, thereby enabling the utilization of the ALLWISE table as a foundation for cross-matching while facilitating multi-processing computations across different celestial regions. 

We selected the ALLWISE catalog as the foundation for several reasons. To begin with, photometry conducted across different epochs yields varying target-detection outcomes. It is imperative to select a reliable input table to mitigate the occurrence of spurious variable sources. Furthermore, the ALLWISE catalog incorporated data from the earliest and most optimal two epochs, resulting in a marginally increased magnitude depth compared to the unTimely catalog. This characteristic enhances the potential for identifying variable sources. Finally, the ALLWISE catalog aligns with the W3 and W4 bands and provides pre-computed cross-match results with other surveys such as 2MASS and \textit{GAIA}, facilitating further investigation of the selected variable sources.

Due to the complex deblending of extended sources in the unTimely catalog, we selected ALLWISE sources under the condition $ext\_flg=0$. We did not constrain other quality flags, as we only needed to ensure the existence of the source in the cross-matching foundation table. For unTimely tables, we required $qf>0.9$ and $nm>4$, ensuring both higher photometric quality and sufficient WISE/NEOWISE exposures contributing to the unWISE coadd at the source position.

We implemented the cross-matching task utilizing STILTS, the command-line tools of the TOPCAT software \citep{2005ASPC..347...29T}, and embedded it in Python code via Jython. Both the ALLWISE and unTimely catalogs divide the full sky into identical astrometric footprints, facilitating multiprocess computation. For each footprint, we cross-matched the ALLWISE table with all unTimely tables at different epochs. This operation was performed independently for the W1 and W2 bands. For the cross-match radius, we adopted 3 arcsec as a balanced choice, minimizing contamination while retaining genuine data points. It is important to note that the radius does not affect the deblending issue, as the central source may lose flux owing to incorrect deblending without altering its position. Following cross-matching, we retained sources with light curves that possessed more than 10 data points to ensure a sufficient length for feature calculation.

\section{variability features} \label{sec:features}

\begin{figure*}[ht!]
\plotone{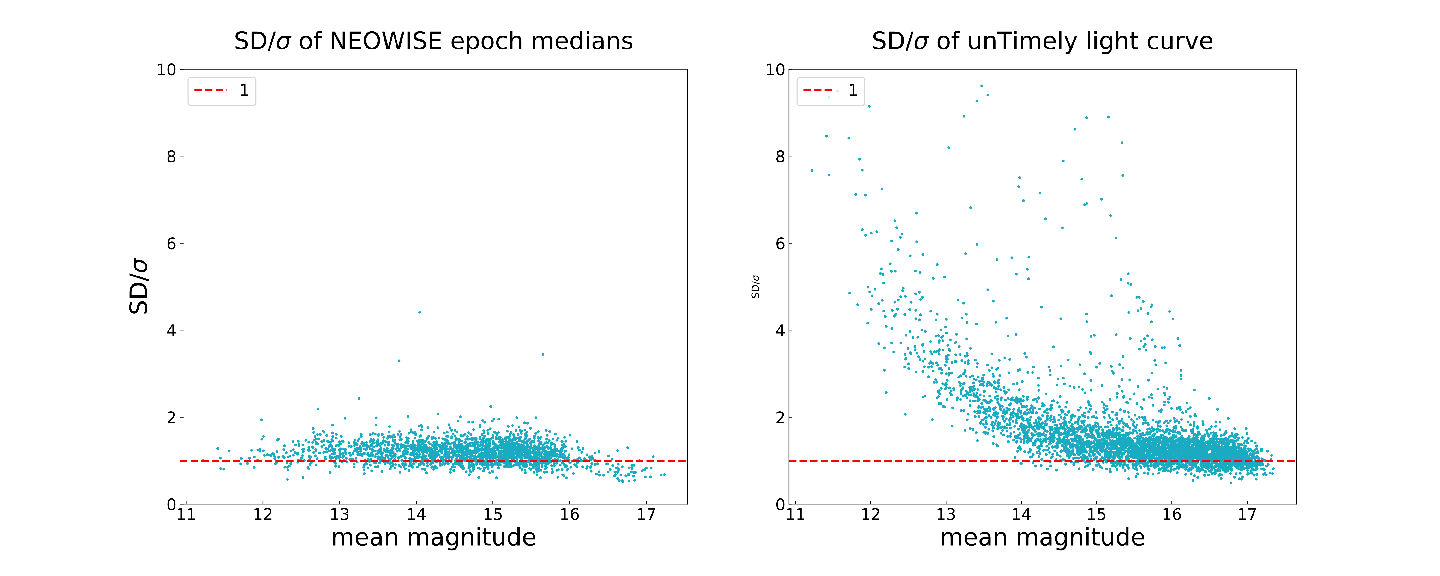}
\caption{The ratio of the standard deviation (SD) to the mean error ($\sigma$) of the NEOWISE epoch medians (left panel) is compared to the unTimely data (right panel), as a function of the mean magnitude. This analysis was based on 5000 randomly sampled SDSS Stripe 82 standard stars. The unTimely data clearly underestimate the photometric uncertainties, particularly for bright sources.
\label{fig:SD_sigma}}
\end{figure*}

To evaluate the variability of the unTimely light curves, we computed 12 features described in \citet{2017MNRAS.464..274S}, along with the mean magnitude. These features are the $\chi^2$ test, reduced $\chi^2$, weighted standard deviation ($\sigma_w$), median absolute deviation (MAD), interquartile range (IQR), robust median statistic (RoMS), normalized excess variance $\sigma^2_{\rm NMS}$, peak-to-peak variability $v$, lag-1 autocorrelatio ($l_1$), Stetson $J$ and $K$ indices, and von Neumann ratio $\eta$. A comprehensive description of these features is provided in \citet{2017MNRAS.464..274S}. Other features described in \citet{2017MNRAS.464..274S} are not applicable to the unTimely data. For example, the consecutive same-sign deviations from the mean magnitude (CSSD) always equal zero because of the limited number of data points. Furthermore, we posit that the uncertainties in FLUX given by the unTimely catalog are significantly underestimated, as the light curve standard deviation of the SDSS Stripe 82 standard stars (considered as non-variable) is considerably higher than the given uncertainties, as shown in Figure \ref{fig:SD_sigma}. Consequently, we omitted
features that require accurate uncertainty estimates \citep{2017MNRAS.464..274S}, and reserve such analyses for future datasets with improved error evaluation.

\begin{figure*}[ht!]
\plotone{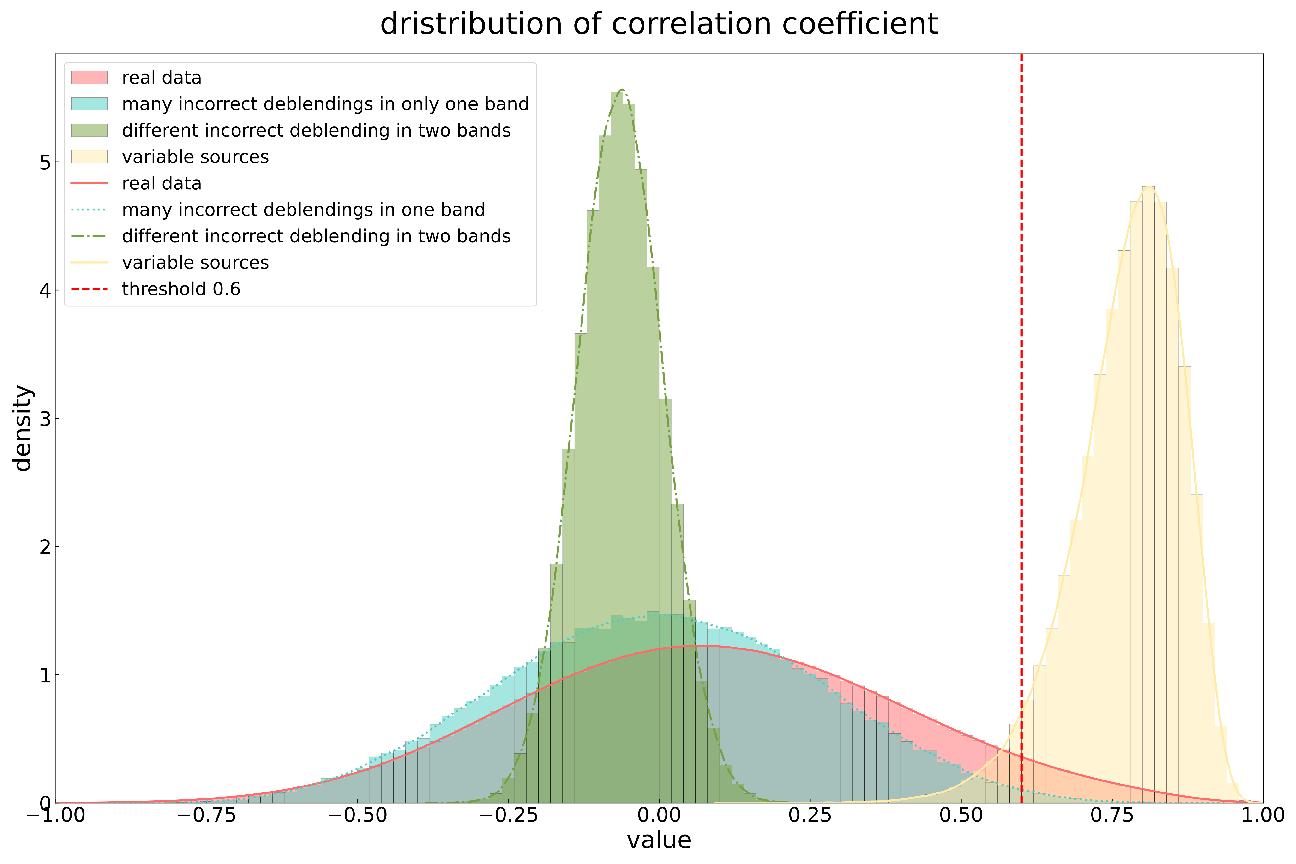}
\caption{Distribution of cross-correlation coefficients derived from a preliminary simulation. The red segment represents the actual data. The blue segment illustrates the distribution when incorrect deblending occurs in only one band. The green segment depicts the distribution when incorrect deblending occurs in both bands at different times. The yellow segment represents the outcome of a preliminary simulation of the variable sources. Notably, the true number of non-variable sources is approximately 100 times greater than that of the variable sources.  
\label{fig:correlation}}
\end{figure*}

Prior to applying the machine learning algorithm on these features, it is essential to eliminate the spurious variable sources resulting from incorrect deblending during the photometry process of the unTimely data. Visual inspection of the light curves shows that incorrect deblending rarely occurs simultaneously in both the W1 and W2 bands. Consequently, a viable approach would be to filter the data using the cross-correlation coefficient of the light curves from both bands. This approach is also beneficial for filtering out certain, non-variable sources. The choice of threshold for the cross-correlation coefficient is partly subjective, reflecting a trade-off between excluding spurious sources and retaining genuine variables. To address this, we conducted a preliminary simulation of untimely variable and non-variable sources, along with potential modes of the artifact. Their cross-correlation coefficients are depicted in Figure \ref{fig:correlation}. Based on this simulation, we selected a relatively lenient threshold of 0.6. Spurious sources that are not excluded will undergo a second filtering process following the selection of variable sources in Section \ref{sec:sample}.

\section{Gaussian mixture model with a Dirichlet process} \label{sec:Gaussian mixture}

Owing to the insufficiency of known invariable sources for use as training data (high-precision photometry is required), we opted for unsupervised algorithms to differentiate between variable and invariable sources. A suitable approach is the Bayesian Gaussian mixture model with a Dirichlet process, a clustering algorithm first introduced into astronomy by \citet{2009MNRAS.400.1897S}. Following the clustering process, the largest cluster is expected to represent non-variable sources. We calculated the Mahalanobis distance to the center of the largest cluster for each source and classified those sources that exceeded the 99.7th percentile of all sources as variables.

Gaussian mixture models (GMMs) are widely employed as clustering models in various disciplines. These models posit that data comprise multiple mixed components, each of which is characterized by a high-dimensional Gaussian probability distribution. The traditional GMM necessitates a prior specification of the number of clusters, which poses challenges in situations in which the determination of the cluster numbers is uncertain. The Bayesian GMM with a Dirichlet process is a non-parametric method designed to address these challenges. This model assumes that the parameters of each Gaussian component follow a distribution sampled from a Dirichlet process.

The Dirichlet Process (DP) is a pivotal Bayesian nonparametric methodology that functions as a prior over distributions, facilitating flexible and adaptive clustering without necessitating the pre-specification of the number of components. It is especially effective in scenarios in which the underlying group structure is either unknown or potentially infinite, such as in the analysis of astronomical time-series data.

In formal terms, a DP is characterized as a stochastic process whose outcomes are probability distributions. Let \( G_0 \) represent a base probability distribution, and \( \alpha \) denote a positive concentration parameter. A random distribution \( G \) is considered to be drawn from a DP:

\begin{equation}
G \sim \text{DP}(\alpha, G_0)
\end{equation}

If, for any finite measurable partition \((A_1, A_2, \ldots, A_K)\) of the sample space, the vector of probabilities \((G(A_1), G(A_2), \ldots, G(A_K))\) follows a Dirichlet distribution:

\begin{equation}
(G(A_1), \dots, G(A_K)) \sim \text{Dir}(\alpha G_0(A_1), \dots, \alpha G_0(A_K))
\end{equation}

DP offers a distribution over distributions, where the base distribution \( G_0 \) signifies the expected value of \( G \), and the parameter \( \alpha \) modulates the variance around \( G_0 \). Elevated values of \( \alpha \) lead to \( G \) being more concentrated around \( G_0 \). A fundamental attribute of DP is its discreteness: with probability one, samples \( G \) drawn from the process are discrete distributions. This characteristic permits the repetition of the same value multiple times, making DP particularly suitable for clustering tasks. Within the context of a mixture model, each data point is assumed to originate from a mixture component parameterized by \( \theta_i \), with these parameters being drawn independently and identically distributed (i.i.d.) from \( G \):

\begin{equation}
\theta_i \mid G \sim G
\end{equation}

Since \( G \) is discrete, there is a positive probability that multiple \( \theta_i \)'s assume the same value, effectively grouping data points into clusters. The ``infinite" aspect arises because DP theoretically supports an unbounded number of clusters; however, in practice, a finite sample will exhibit a finite number. This renders the infinite GMM a natural application of DP. In such a model, the number of Gaussian components does not need to be specified in advance; instead, it is inferred from the data. The full generative process can be summarized as follows:

\begin{equation}
G \sim \text{DP}(\alpha, G_0)
\end{equation}
\begin{equation}
\mu_k, \Sigma_k \sim G
\end{equation}
\begin{equation}
\mathbf{x}_i \mid (\mu_k, \Sigma_k) \sim \mathcal{N}(\mu_k, \Sigma_k)
\end{equation}

Posterior inference in DP mixture models is frequently conducted using Markov Chain Monte Carlo (MCMC) techniques, such as the Gibbs sampler. In our implementation, we employed variational inference (VI) to enhance computational scalability, utilizing the capabilities of the \texttt{scikit-learn} library. This methodology enables us to efficiently approximate the posterior distribution of cluster assignments and parameters, thereby providing a probabilistic basis for identifying groups, such as variable and non-variable stars in astronomical surveys, in a fully data-driven manner. The application of the DP prior thus presents a principled, flexible, and robust framework for unsupervised learning, wherein the model complexity is adaptively aligned with the observed data.

\section{The variable sources sample} \label{sec:sample}
\begin{figure*}[ht!]
\plotone{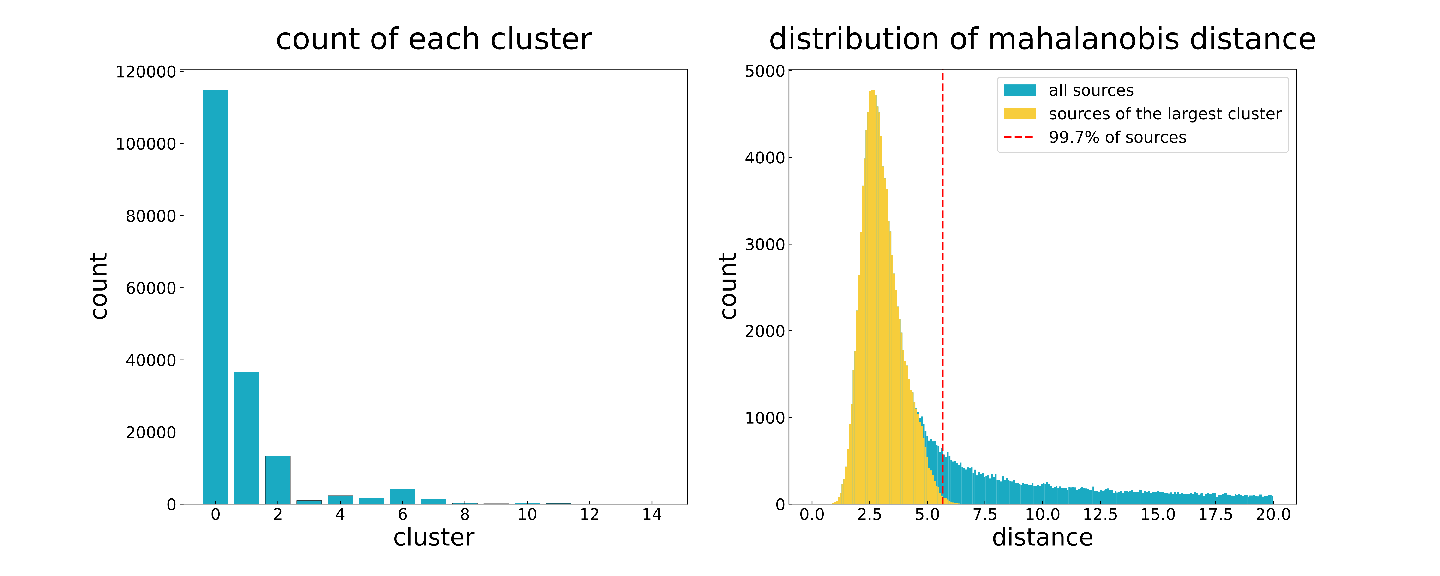}
\caption{The left panel illustrates the distribution of the cluster results derived from the algorithm applied to the 14.5-14.6 subset. The largest cluster corresponded to non-variable sources. The right panel depicts the Mahalanobis distance of the sources from the center of the largest cluster, highlighting a threshold at the 99.7th percentile.  
\label{fig:example_bin}}
\end{figure*}

Given that the values of features are associated with the magnitude of sources, we performed clustering and distance selection operations on 0.1 magnitude-binned subsets of light curve data. To ensure that each subset contained a sufficient number of data items for the clustering algorithm (at least several hundred), we determined the valid magnitude ranges to be 8.0–18.3 for the W1 band and 8.0–17.2 for the W2 band. Figure \ref{fig:example_bin} presents a representative subset of the W1 band, ranging from 14.5 to 14.6. It illustrates both the histogram of clusters and the distribution of the Mahalanobis distance in relation to the center of the largest cluster, emphasizing a 99.7\% distance range.

\begin{figure*}[ht!]
\plotone{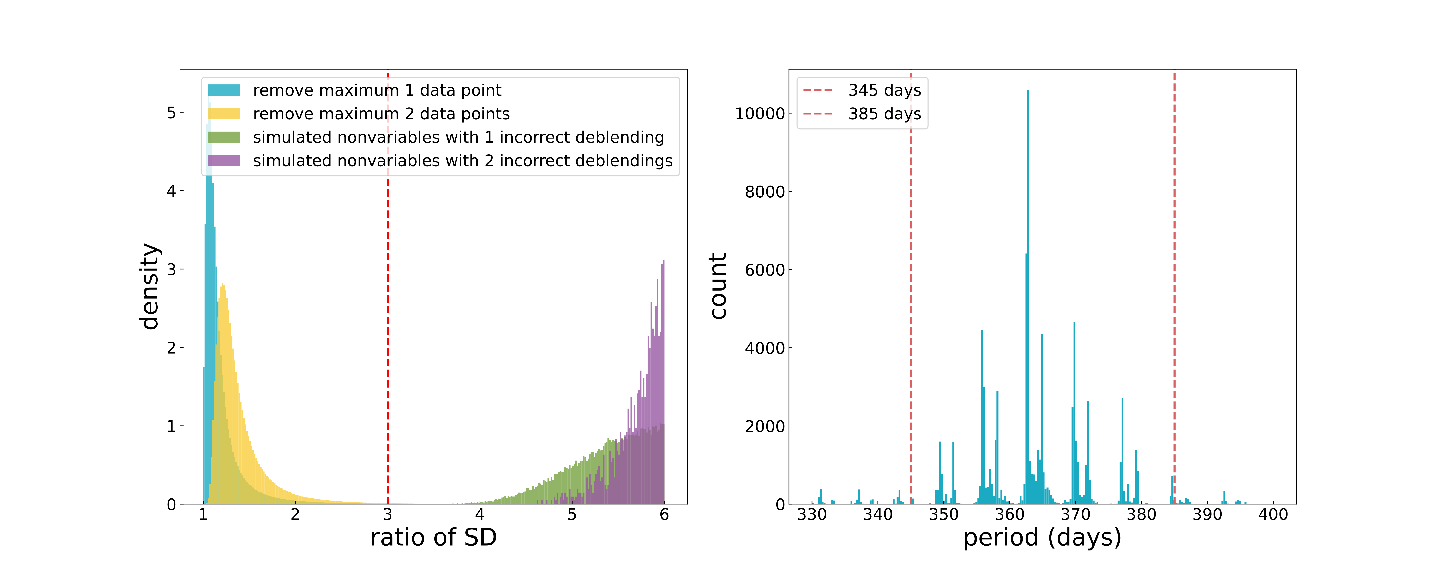}
\caption{The criteria for filtering out the two types of artifacts. The left panel displays the ratio of the standard deviation before and after the removal of the largest one or two data points, incorporating both the actual data and a rough simulation. The right panel illustrates the distribution of the Lomb-Scargle period around one year.   
\label{fig:filtering}}
\end{figure*}

\begin{figure*}[ht!]
\plotone{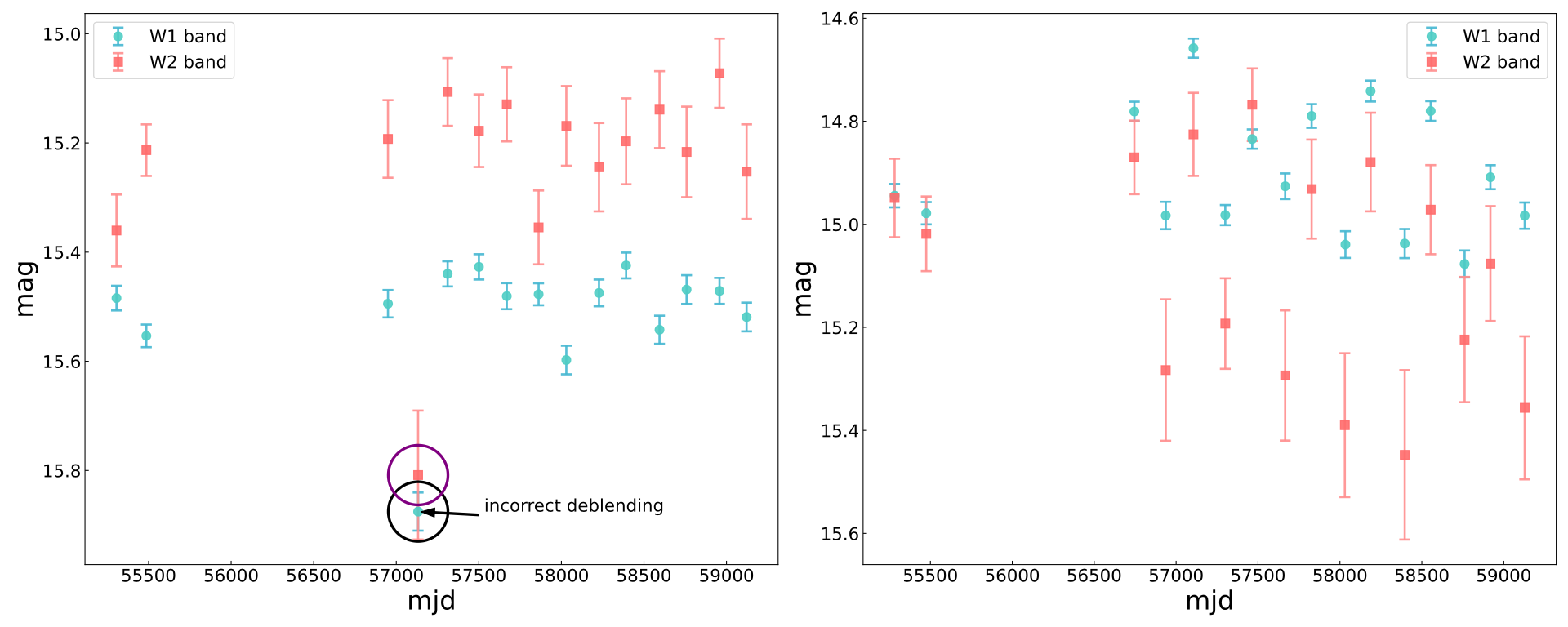}
\caption{Two primary types of artificial variable light curves. The left panel illustrates the artifact resulting from incorrect deblending, while the right panel depicts the artifact caused by the WISE satellite instrument.  
\label{fig:articicial_light_curve}}
\end{figure*}

Following the clustering and selection operations, certain artifacts persist, contaminating the variable samples, primarily belonging to two categories. The first category comprises residual spurious sources resulting from incorrect deblending, which are not eliminated by the previous cross-correlation coefficients technique. These sources typically exhibit one or two erroneous deblending data points. A typical instance is shown in Figure \ref{fig:articicial_light_curve}. To address this, we employed the ratio of the standard deviation before and after removing the largest one and two magnitude data points, as incorrect deblending generally leads to a decrease in flux. We have determined a threshold of 3, as illustrated in Figure \ref{fig:filtering}. The second category, resulting from an issue with the WISE satellite instrument, is affected by the latent image artifact from a nearby bright source that appears in the preceding image in the scan \citep{2021ApJS..252...32J}. In a specific region, the WISE scan direction on the sky reverses every six months due to the sun-synchronous orbit, leading to an artificial six-month period, as illustrated in Figure \ref{fig:articicial_light_curve}. These sources can be filtered out using the Lomb-Scargle periodogram around one year; we excluded sources with a most likely period ranging from 345 days to 385 days, as shown in Figure \ref{fig:filtering}. It is important to note that this criterion was designed to ensure the purity of the final sample. A significant portion of the excluded sources may still represent valid variable sources. Therefore, we will provide an exclusion table in addition to the main variable catalog.

Following the aforementioned procedure, we obtained a sample of variable sources, with 8,256,042 sources identified in the W1 band and 7,147,661 in the W2 band. Of these, 4,294,039 sources exhibited variability in both bands. We conducted a visual inspection of thousands of randomly sampled light curves from these sources and effectively confirmed their variability.

\section{Outliers and special varibility} \label{sec:Outliers}

To identify potential sources of scientific significance, we initially computed the Lomb-Scargle period and quasi-periodicity ($Q$) and flux asymmetry ($M$) statistics \citep{2014AJ....147...82C} for each light curve. These parameters were incorporated into the final variable source catalogue to assist users in locating sources of interest.

To advance our analysis, we employed algorithms to identify outliers in terms of light curve morphology. Initially, we established a parameter to identify sources with valid morphology, namely those whose variability was not excessively stochastic. We define a parameter termed ``smoothness", which is calculated as one minus the ratio of the number of sign changes in the residuals of a light curve's magnitude sequence to the total length of the light curve.

A light curve with a smoothness value of 1 indicates that it is either monotonically increasing or decreasing. Sources with a smoothness greater than 0.5 were utilized in subsequent analyses, including feature extraction and anomaly detection.

\begin{figure*}[ht!]
\plotone{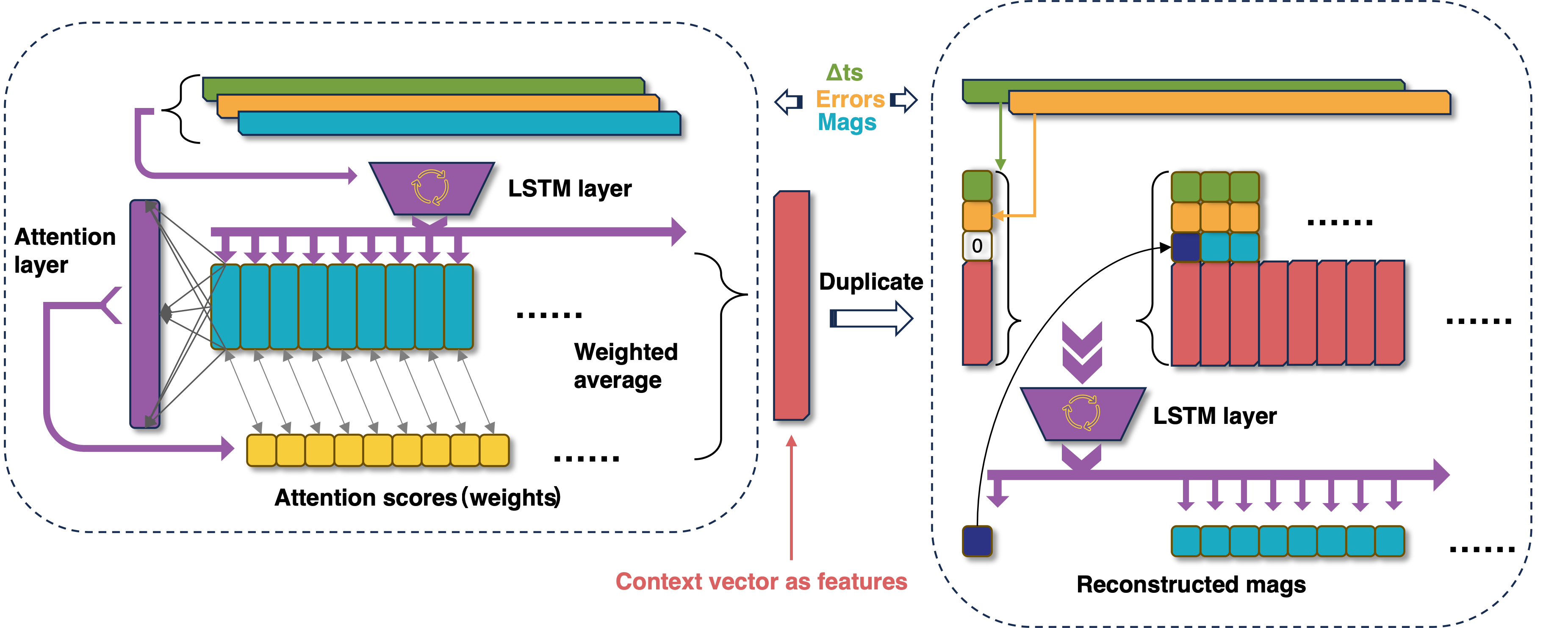}
\caption{The architecture of the autoencoder neural network to extract morphological features of light curve.
\label{fig:network_structure}}
\end{figure*}

As the variability features discussed in Section \ref{sec:features} are solely intended to ascertain whether a source is variable, developing new features to characterize the morphology is necessary. In this context, we employed an autoencoder neural network for feature extraction. The network was designed to reconstruct the magnitude sequence of the light curves based on the given time and error sequences. The network architecture is depicted in Figure \ref{fig:network_structure}. The temporal data of a light curve were transformed into a residual form, serving as input alongside the magnitudes and errors. The encoder comprises a Long Short-Term Memory (LSTM) recurrent layer and a content-based attention layer, which converts the LSTM output sequence into a fixed-length context vector. The context vector is the desired feature vector. The decoder, also an LSTM, receives as input the concatenation of the context vector, time residual, and predicted magnitude from the previous step. The latter is generated by two fully connected layers that transform the LSTM output vector into a magnitude. To enhance the reconstruction accuracy, we developed a modified mean squared error (MSE) loss function that assigns a weight of 1.5 times to the extreme points of a light curve.

\begin{figure*}[ht!]
\plotone{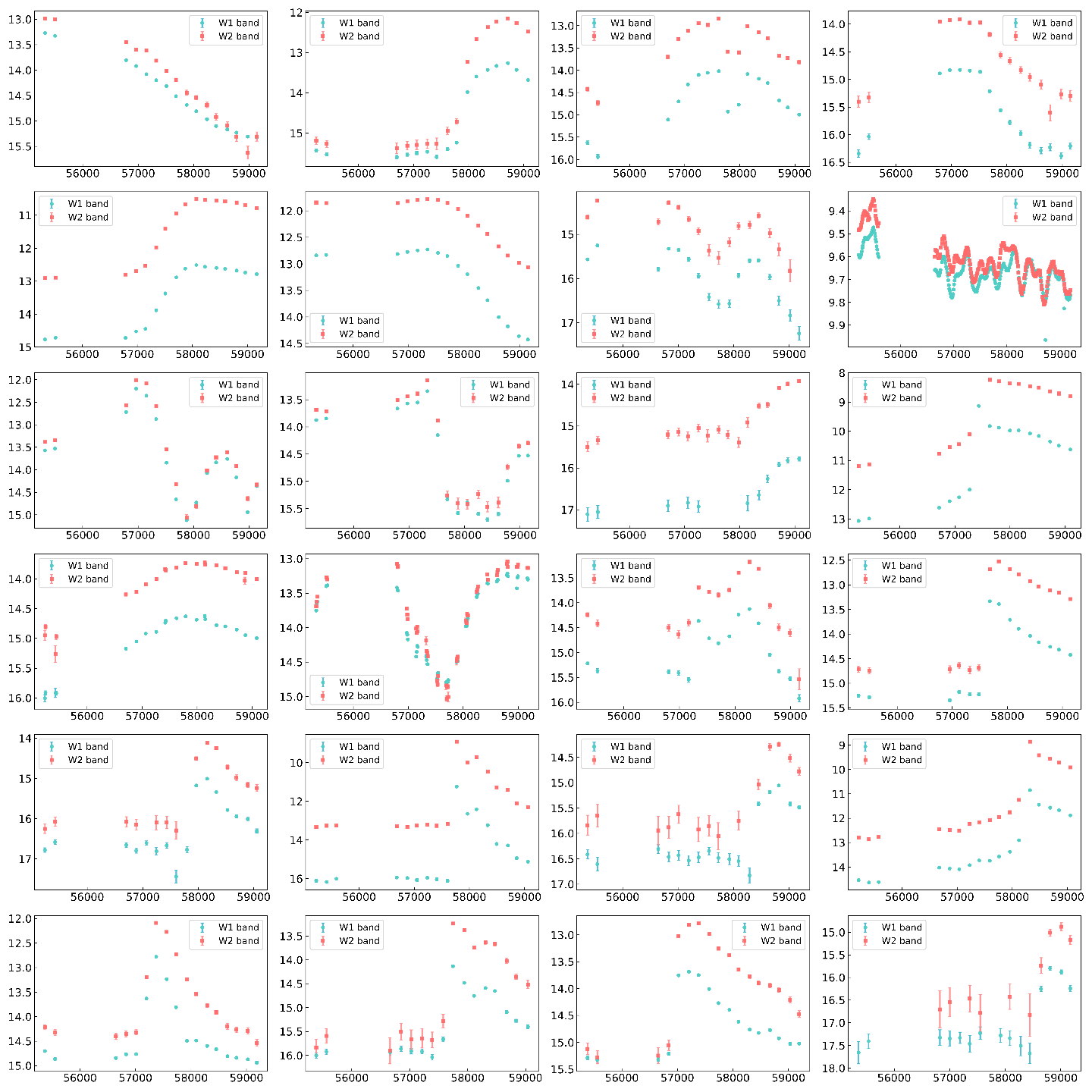}
\caption{Some example light curves for the identified outliers.
\label{fig:special_light_curves}}
\end{figure*}

Following the feature extraction process, two outlier detection algorithms were employed to identify unique sources: the Isolation Forest (iForest) and the Local Outlier Factor. The iForest model, as introduced by \citet{4781136}, is an efficient algorithm designed to isolate instances based on the premise that anomalies are both infrequent and distinct, thus requiring fewer random partitions within a tree structure. This model is particularly adept at identifying global anomalies. Conversely, the Local Outlier Factor algorithm proposed by \citet{10.1145/335191.335388} operates on a density-based principle. It identifies outliers by comparing the local density of a point to the densities of its neighboring points, making it highly effective for detecting local anomalies within complex data structures. The implementation of these two algorithms will facilitate the identification of both global and local anomalies. The anomaly scores generated by both algorithms were incorporated into our final catalog, as exemplified in Table \ref{tab:description}. Figure \ref{fig:special_light_curves} illustrates some of the identified distinctive light curves.

\begin{deluxetable*}{rl}
\tablewidth{0pt}
\tablecaption{Description of table columns of our variable source catalogue} \label{tab:description}
\tablehead{
\colhead{Column Name} & \colhead{Explanation}
}
\startdata
RA & right ascension \\
DEC & declination \\
P & Lomb-scargle best period \\
FAP & false alarm probability of Lomb-scargle best period \\
Q & quasi-periodicity \\
M & flux asymmetry \\
amplitude & the maximum magnitude minus the minimum magnitude of a light curve \\
smoothness & defined in Section \ref{sec:Outliers} \\
iForest score & anomaly score from Isolation Forest algorithm \\
LOF score & anomaly score from Local Outlier Factor algorithm \\
\enddata
\tablecomments{There are two tables, the main table and the excluded table. The tables will be available online soon}
\end{deluxetable*}

\section{Discussion} \label{sec:discussion}

Owing to the unprecedented comprehensiveness and depth of our catalogue, several potential scientific directions warrant investigation:
\begin{enumerate}
\item Investigate long-period variable stars, such as Mira variables and semi-regular variable stars. These stars typically exhibit variability on an annual timescale, allowing for the measurement of their amplitude and period or quasi-periodicity through our catalog.

\item The search and classification of young stellar objects (YSOs) are of significant interest, particularly those exhibiting notable variability in the mid-infrared band. Certain types of YSOs, especially those experiencing bursts or deep dips, demonstrate this kind of variability. These objects can be effectively identified and measured using our catalogue, which is cross-referenced with photometry catalogues from other bands.

\item Our catalog, which operates in the mid-infrared band and encompasses the entire sky, facilitates the search for luminous red novae (LRNe) and supernovae. This comprehensive coverage can aid in identifying LRNe and certain types of long-term supernovae that may have been overlooked by other sky survey projects over the past decade due to incomplete sky coverage.

\item Investigate the phenomena of changing-look and changing-state active galactic nuclei (AGNs) as well as supermassive black hole binaries. Some potential sources of these unique types have not been adequately identified by current spectral sky surveys, which are unable to confirm their status as AGNs. Our catalog facilitates the initial identification of candidates, followed by additional observations to confirm their classification.

\item The search for obscured AGNs is challenging because of the presence of substantial dust, which renders them difficult to detect in the optical band. However, these AGNs may exhibit variability in the mid-infrared band. Our catalog facilitates the identification of such AGNs by highlighting those that demonstrate relatively significant and prolonged variability.

\item It is important to note that although unTimely data possess a deeper limiting magnitude than NEOWISE single exposure (L1b) data available on IRSA, most sources that are not exceedingly faint can obtain more detailed information through the single exposure data. Therefore, an efficient approach to utilizing our catalog is to initially identify candidates and subsequently refer to the L1b data to acquire specific information for each epoch.
\end{enumerate}

As the WISE satellite ends its life in a trail of light through the atmosphere, it is imperative to develop a comprehensive catalog of its variable sources. Our current study initiates this process, albeit with certain limitations. Due to the unTimely catalog's exclusion of complete data from the final stages of WISE's operational life, some sources exhibiting recent variability may be overlooked. Additionally, the photometric methods employed by the unTimely catalog, which result in a trade-off between completeness and purity due to incorrect deblending, may also lead to the omission of some variables. Future endeavors to achieve a comprehensive catalog of WISE mid-infrared variable sources will enable systematic research into the statistical distribution of mid-infrared variables across the entire sky.

Furthermore, our methodology is applicable to the original NEOWISE single exposure (L1b) data. The variability index in the ALLWISE table only approximately accounts for variability during the first half-year of observation, resulting in the omission of a systematic search for short-term variable sources. This task may be undertaken by individuals who possess the necessary permission to access the entire database.

\section{Conclusion}
We have constructed the largest mid-infrared variability catalog to date, containing millions of variable sources identified from the unTimely database. This work demonstrates that even with relatively sparse temporal sampling and modest photometric accuracy, mid-infrared time-domain data can be transformed into a powerful resource for variability studies across the sky. The catalog not only enables systematic statistical analyses of variable stars, AGNs, and young stellar objects, but also provides a fertile ground for the discovery of rare and extreme phenomena that are inaccessible in optical bands due to dust obscuration.
Looking ahead, this dataset will serve as a critical bridge between current optical time-domain surveys and future multi-wavelength facilities. By cross-matching with \textit{Gaia}, LSST, \textit{Roman}, and other upcoming large-scale surveys, the catalog will enable multi-band variability studies with unprecedented depth and coverage. Moreover, the identification of extreme outliers opens new opportunities for targeted follow-up observations with space- and ground-based telescopes, offering insights into the physics of accretion, stellar explosions, and dust-enshrouded environments. Ultimately, this work establishes mid-infrared time-domain astronomy as a vital component of the next decade of astrophysical research, laying the groundwork for advancing our understanding of both Galactic and extragalactic variable populations.

\begin{acknowledgments}
\justifying
This work is supported by the National Natural Science Foundation of China (NSFC; grant Nos.12273076, 12203077, 12373110, 12133001), the China Manned Space Program with grant no. CMS-CSST-2025-A06 and Hebei Natural Science Foundation (A2024205020). This work makes use of data products from the Wide ﬁeld Infrared Survey Explorer, which is a joint project of the University of California, Los Angeles, and the Jet Propulsion Laboratory/California Institute of Technology, funded by the National Aeronautics and Space Administration.

\end{acknowledgments}

\bibliography{sample7}{}
\bibliographystyle{aasjournalv7}



\end{document}